\documentclass[11pt]{article}
\usepackage{mathptmx}
\usepackage{sprinconf}
\usepackage{graphicx}
\usepackage{bm}\usepackage{amssymb,amsmath}

\usepackage{times}\usepackage{dsfont}
\newcommand{\ei}{\end{itemize}}
\newcommand{\be}{\begin{eqnarray}}
\newcommand{\ee}{\end{eqnarray}}
\newcommand{\ba}{\begin{array}}
\newcommand{\ea}{\end{array}}
\newcommand{\bc}{\begin{center}}
\newcommand{\ec}{\end{center}}
\newcommand{\bt}{\begin{tabular}}
\newcommand{\btab}{\begin{table}}
\newcommand{\et}{\end{tabular}}

\begin{document}
\sloppy \raggedbottom
\setcounter{page}{1}
\setcounter{figure}{0}
\setcounter{equation}{0}
\setcounter{footnote}{0}
\setcounter{table}{0}
\setcounter{section}{0}

\title{Modular Invariants and Fischer-Griess Monster
\footnote{Lecture given by M. Jankiewicz at 26th International Colloquium on Group Theoretical Methods in Physics, June 26- June 30 2006, The Graduate Center of the City University of New York.}
}
\author{Marcin Jankiewicz\dag, Thomas W. Kephart\dag} 
\affil{\dag\ Department of Physics and Astronomy, Vanderbilt
University, Nashville, TN 37235, USA}

\vspace{-25pt} 

\beginabstract
\noindent
We show interesting relations between extremal partition functions of a family of conformal field theories and dimensions of the irreducible representations of the Fischer-Griess Monster sporadic group. We argue that these relations can be interpreted as an extension of Monster moonshine.
\endabstract

\vspace{-25pt} 


\section{Introduction}
Partition functions of conformal field theories with central charge $c=24\cdot k$, where $k=1,2,...$,  are modular invariants. There exists a relation between CFT with $c=24$, i.e. $k=1$, and the Monster sporadic group. This is the famous Monster moonshine theorem that states that coefficients of the $q$-expansion are related in a simple manner to the dimensions of the irreducible representations of the this largest sporadic group. Here, we are going conjecture a natural extension of this relation to the higher dimensional, i.e. $k>1$, case. 
We are going to use lattices, or more precisely their $\Theta$-functions to describe a given CFT. 
The $q$-expansion of a $\Theta$-function of a lattice $\Lambda$ is given as
\be Z_{\Lambda}=\sum_{x\in\Lambda}N(m)q^{m}\,,\ee
where we sum over all vectors $x$, in the lattice $\Lambda$, with length $m=x\cdot x$. $N(m)$ 
is the number of vectors of norm $m$ and $q\equiv e^{i\pi\tau}$ where $\tau$ is the modular parameter.
The spectra of meromorphic conformal field theories can be expressed in terms of partition functions of even self-dual lattices.
This means that the exponent $m$ in the $q$-expansion will be necessarily an even number. 
Both self-duality and evenness of a lattice correspond to invariance of a partition function $\mathcal{Z}$ (which is closely related to $Z_{\Lambda}$) under the generators $S$ and $T$ of a modular group $SL(2,\mathds{Z})$.

Formally, $Z_{\Lambda}$ of a $d$ dimensional lattice $\Lambda$ is a modular form of weight $d/2$. The
partition function $\mathcal{Z}$ of a lattice is defined as follows
\be\label{partdef}\mathcal{Z}=Z_{\Lambda}/\eta^{d/2}\,,\ee 
where $\eta(q)=q^{1/12}\prod_{m=1}^{\infty}(1-q^{2m})$
is Dedekind $\eta$-function which is a modular form of weight $1/2$.
The partition functions of all the 24 dimensional even self-dual lattices (the Niemeier lattices) 
can be written as
\be\label{part1}\mathcal{Z}=\left[J+24(h+1)\right]\eta^{24}\,,\ee 
where $h$ is the Coxeter number of a given lattice. For example, $h=0$ corresponds to the famous 
Leech lattice, $h=30$ to the Niemeier lattice based on a root system of $E_{8}^{3}$, etc. 
Physically $24(h+1)$ corresponds to a number of massless states in a given theory.
Using a technique presented in \cite{Jankiewicz:2005rx,Jankiewicz:2005wh}, one can choose any Niemeier lattice $\Lambda_{1}$ 
to generate the $\Theta$-function of another Niemeier lattice $\Lambda_{2}$. In \cite{Jankiewicz:2005rx} it was shown 
that it is possible to generate extremal partition functions  \cite{Sloaneweb}
by taking the $k^{th}$ power of (\ref{part1}) and treating $x_{i}=24(h_{i}+1)$ (where $i=1,...,k$) as a free parameter.


\section{CFTs with $c=24\cdot k$: Systematic Approach}
Different choices of constant parameters $x_{i}s$ correspond to 
different $\Theta$-functions. One can find corresponding extremal 
partition functions, (\ref{partdef}) and write them as $q$-expansions of the form
\be\prod_{i=1}^{k}(J+24+x_{i})=\frac{1}{q^{2k}}\left[1+\sum_{m=(k-1)}^{\infty}f_{2m}(x_{1},...,x_{k})q^{2m-2k}\right]\ee
As an example, we will work with a choice of $k-1$ parameters $x_{i}$s that eliminates coefficients of all but one term with negative powers in the 
$q$-expansion that correspond (in a field theoretic language) to tachyonic states. In this setup we are left with only one free parameter $x_{k}$. 
Different choices of $x_{k}$ would correspond to different partition function candidates for conformal field 
theories with $c=24\cdot k$. Here we list the first three cases:
\begin{subequations}
\begin{align}
&\mathcal{G}_{1}(x_{1})\!=\!\frac{1}{q^{2}}+(24+x_{1})+196884q^{2}+...\\
&\mathcal{G}_{2}(x_{2})\!=\!\frac{1}{q^{4}}+(393192-48x_{2}-x_{2}^{2})+42987520q^{2}+...\\ \label{tach}
&\mathcal{G}_{3}(x_{3})\!=\!\frac{1}{q^{6}}+(50319456-588924x_{3}+72x_{3}^{2}+x_{3}^{3})+2592899910q^{2}+...
\end{align}\end{subequations}
Notice that in each case all of the tachyonic states (except the lowest one) are absent. Since the allowed 
values \cite{Harvey:1988ur} of the coefficient of $q^{0}$ are integers that run from zero to the value of the 
$q^{2}$ coefficient, one can easily find the number of ``allowed'' partition functions in $24k$ dimensions \cite{Jankiewicz:2005rx}.\\
\section{Monster Moonshine and its Extension}
The extremal 24 dimensional case has been shown to be related to the Fischer-Griess monster group. 
In mathematics this fact is known as Monster moonshine (\cite{Dolan:1989kf} and \cite{Borch}). 
One can evaluate $\mathcal{G}_{1}$ at $x_{1}=-24$ which corresponds to the $j$-invariant to find
\begin{eqnarray}j=\frac{1}{q^{2}}+196884q^{2}+ 21493760q^{4} + 864299970q^{6} + 20245856256q^{8}+...\,.\end{eqnarray}
\noindent
The coefficients of this expansion decompose into dimensions of the irreducible representations of the 
Monster\footnote{for explicit realization of the Monster moonshine see \cite{Jankiewicz:2005rx}.}, where we use the notation $j=\frac{1}{q^{2}}+j_{2}q^{2}+j_{4}q^{4}+...\,$.
Following this interpretation of the Monster, one can easily generate moonshine in 
 higher dimensional cases, i.e., one can express coefficients of any partition functions, 
for example $\mathcal{G}_{k}(x_{k})$, for any choice of $k$, in terms of the dimensions of 
irreducible representations of the Monster group. We present a few of our results \cite{Jankiewicz:2005rx} in Table-\ref{tab-mon3}, 
where coefficients of $G_{k}(x_{k})$ are expressed in terms of the coefficients of the invariant 
function $j$, 
that (via the original Monster moonshine) are related to the Monster.
We notice \cite{Jankiewicz:2005rx} that the coefficients $g_{2n}$ fall into patterns with period $k!$,
and conjecture that this periodicity also continues to hold for all $k$.
The polynomial conditions to be satisfied to find the
extremal partition functions for large $k$ become increasingly more
difficult to solve with increasing $k$, so we do not have results
for $k>6$.\\
Table-\ref{tab-mon3} give the general periodicity in coefficients of $\mathcal{G}_{k}(x_{k})$ for $k < 6$. 
\begin{table}
{\scriptsize\begin{tabular}{|l|l||l|l|}
\hline
$k=2$        & $k=2$                               & $k=3$        & $k=3$                     \\
\hline
$g_{4i+2}$   & $2j_{2(4i+2)}$                      & $g_{6i+2}$   & $3j_{3(6i+2)}$            \\
$g_{4i+4}$   & $2j_{2(4i+4)}+j_{2(2i+2)}$          & $g_{6i+4}$   & $3j_{3(6i+4)}$            \\
             &                                     & $g_{6i+6}$   & $3j_{3(6i+6)}+j_{2i+2}$   \\
\hline
$k=4$        & $k=4$                               & $k=5$        & $k=5$                     \\
\hline
$g_{8i+2}$   & $4j_{4(8i+2)}$                      & $g_{10i+2}$  & $5j_{5(10i+2)}$           \\
$g_{8i+4}$   & $4j_{4(8i+4)}+2j_{2(2i+4)}$         & $g_{10i+4}$  & $5j_{5(10i+4)}$           \\
$g_{8i+6}$   & $4j_{4(8i+6)}$                      & $g_{10i+6}$  & $5j_{5(10i+6)}$           \\
$g_{8i+8}$   & $4j_{4(8i+8)}+2j_{(8i+8)}+j_{2i+2}$ & $g_{10i+8}$  & $5j_{5(10i+8)}$           \\
             &                                     & $g_{10i+10}$ & $5j_{5(10i+10)}+j_{2i+2}$ \\
\hline
 \end{tabular}}
\\
\caption{Periodicity of the coefficients $g_{n}$ for $c=24\cdot k$ extremal partition functions $\mathcal{G}_{k}$ in terms of coefficients  $j_{2n}$ of the modular function $j$.}
\label{tab-mon3}
\end{table}
To summarize, when $k=1$ it is known via standard Monster Moonshine that the coefficients of $j$ decompose into Monster representations \cite{Borch}. The fact that all the higher $k$ coefficients also decompose into Monster representations indicates that they have large symmetries containing the Monster and the fact that they have these symmetries may  indicate that they are related to $24k$ dimensional lattices.

\section{Final Comments}

We have shown that there exists a family of transformations, more general then of simple $\mathds{Z}_{2}$ form, that relates the members of the class of holomorphic conformal field theories, i.e., the Niemeier lattices. Furthermore these transformations connect non-Niemeier modular invariant $c=24$ $\Theta$-functions. These results generalize to any $c=24\cdot k$ case (in particular, when the resulting parametrization corresponds to $24k$ dimensional lattice).
In $24$ dimensions there is one extremal partition function, corresponding
to a Leech lattice, whose $q$-expansion coefficients can be written in
the form of a linear combinations of dimensions of the irreducible
representations of the Monster group; this relationship is at the core of the Monster Moonshine. 
The decomposition is possible only in the
extremal case, since in all other cases a non-zero constant term in the $q$-expansion would be present. The presence of this term in the expansion
would imply the introduction of an enormous unnatural set of singlets in
the decomposition. We believe a similar situation occurs at $24k$
where we have extended this argument. Instead of a
Leech lattice we have to deal with higher dimensional extremal
partition functions (but note \cite{Jankiewicz:2005rx} that there is more than one type of extremal partition function in 24k dimensions). Existence of $24k$
extremal lattices
for $k>2$ is only a conjecture \cite{Sloaneweb}, but our results
are consistent with and provides supporting evidence for this
conjecture.
One can use our generalized version of Monster
Moonshine to postulate that we already have the $q$-expansion of higher
dimensional extremal lattices, and that the symmetry provided by the
Monster decomposition can be used to learn more about these lattices.



We suggest that the extremal $\cal{G}$ partition functions will generate new $c=24\cdot k$ CFTs. We know the first at $k=1$ corresponds to the Leech lattice, the second $k=2$ case also corresponds to the known lattice, $P_{48}$, and since the higher $k$ cases all possess Monster symmetry we conjecture that it is likely that they correspond to CFTs constructed on extremal lattices in
$24k$-dimensions. To the best of our knowledge, Monster symmetry was not known to come into play except at $k=1$.

Using the techniques presented in \cite{Jankiewicz:2005rx}, one can construct a large class of conformal field theories with central charge that is a multiple of 24. We have demonstrated (or at least conjectured) the possibility of  a new realization of Monster moonshine. This is realized as a periodicity in a pattern of coefficients in $q$-expansions of the extremal partition functions.

\section*{Acknowledgments}
This work was supported in part by U.S. DoE grant \#~DE-FG05-85ER40226.




\end{document}